\documentclass[journal=jacsat,manuscript=perspective]{achemso}

\usepackage[version=3]{mhchem} % Formula subscripts using \ce{}
\usepackage[T1]{fontenc}       % Use modern font encodings

\usepackage{subfig}
\author{Shiyang Long}
\affiliation[School of Life Sciences, Jilin University, Changchun, China 130012]
{School of Life Sciences, Jilin University, Changchun, China 130012}
\author{Pu Tian}
\email{tianpu@jlu.edu.cn}
\phone{+86 (0)431 85155287}
\affiliation[School of Life Sciences, Jilin University, Changchun, China 130012]
{School of Life Sciences, Jilin University, Changchun, China 130012}
\alsoaffiliation[MOE Key Laboratory of Molecular Enzymology and Engineering, Jilin University, Changchun, China 130012]
{MOE Key Laboratory of Molecular Enzymology and Engineering, Jilin University, Changchun, China 130012}

\title[]{Configurational space continuity and free energy calculations}

\abbreviations{}

\begin{document}

%%%%%%%%%%%%%%%%%%%%%%%%%%%%%%%%%%%%%%%%%%%%%%%%%%%%%%%%%%%%%%%%%%%%%
%% The "tocentry" environment can be used to create an entry for the
%% graphical table of contents. It is given here as some journals
%% require that it is printed as part of the abstract page. It will
%% be automatically moved as appropriate.
%%%%%%%%%%%%%%%%%%%%%%%%%%%%%%%%%%%%%%%%%%%%%%%%%%%%%%%%%%%%%%%%%%%%%
%\begin{tocentry}
%
%Some journals require a graphical entry for the Table of Contents.
%This should be laid out ``print ready'' so that the sizing of the
%text is correct.
%
%Inside the \texttt{tocentry} environment, the font used is Helvetica
%8\,pt, as required by \emph{Journal of the American Chemical
%Society}.
%
%The surrounding frame is 9\,cm by 3.5\,cm, which is the maximum
%permitted for  \emph{Journal of the American Chemical Society}
%graphical table of content entries. The box will not resize if the
%content is too big: instead it will overflow the edge of the box.
%
%This box and the associated title will always be printed on a
%separate page at the end of the document.
%
%\end{tocentry}

%%%%%%%%%%%%%%%%%%%%%%%%%%%%%%%%%%%%%%%%%%%%%%%%%%%%%%%%%%%%%%%%%%%%%
%% The abstract environment will automatically gobble the contents
%% if an abstract is not used by the target journal.
%%%%%%%%%%%%%%%%%%%%%%%%%%%%%%%%%%%%%%%%%%%%%%%%%%%%%%%%%%%%%%%%%%%%%
\begin{abstract}
Free energy is arguably the most important function(al) for understanding of molecular systems. A number of rigorous and approximate methods, and numerous scoring functions for free energy calculation/estimation have been developed over a few decades. However, the continuity of an macrostate (or path) in configurational space has not been well articulated. In this perspective, we discuss the relevance of configurational space continuity in development of more efficient and reliable next generation free energy methodologies.   
\end{abstract}

%%%%%%%%%%%%%%%%%%%%%%%%%%%%%%%%%%%%%%%%%%%%%%%%%%%%%%%%%%%%%%%%%%%%%
%% Start the main part of the manuscript here.
%%%%%%%%%%%%%%%%%%%%%%%%%%%%%%%%%%%%%%%%%%%%%%%%%%%%%%%%%%%%%%%%%%%%%
\section{Introduction}
\begin{figure}
\centering 
{\includegraphics[width=6.5in]{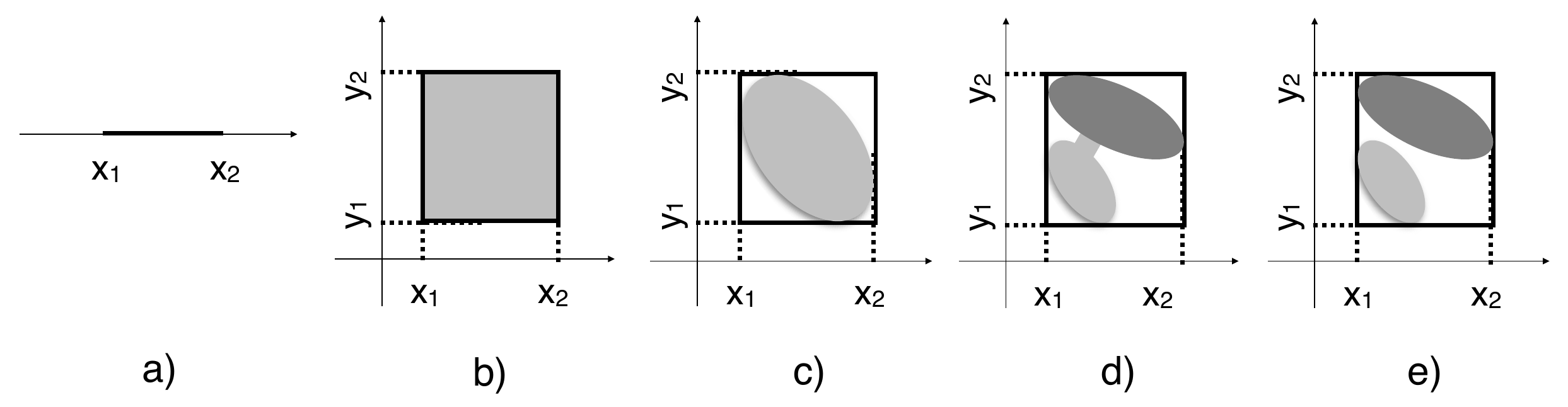}}
\caption{Schematic representation of continuity for domain of simple functions and functionals. a) a domain defined by $[x_1, x_2]$ on one dimension, b) a domain defined by $[x_1,x_2]$ and $[y_1,y_2]$ in two dimensions with $x$ and $y$ being independent, c) a continuous domain defined by $[x_1,x_2]$ and $[y_1,y_2]$ in two dimensions with $x$ and $y$ being correlated, d) a quasi-continuous domain defined by $[x_1,x_2]$ and $[y_1,y_2]$ in two dimensions with $x$ and $y$ being correlated, e) two discrete domains defined by $[x_1,x_2]$ and $[y_1,y_2]$ in two dimensions with $x$ and $y$ being correlated.} 
\label{fig:dofcorr}
\end{figure}

Despite the fact that we live in a three dimensional world, our intuitive comfortable zone is, unfortunately, limited to one dimension. We may figure out objects and properties in two and three dimensions quite well with some training and aid of visualization tools, which becomes difficult beyond three dimensions. In carrying out computational analysis of physical systems, continuity is a highly important property. There are two aspects of continuity, namely that of a given function itself and that of the domain of a given function, and we focus on the later in this perspective. In one dimension, continuity of the domain of a function is trivial as a line segment is always continuous as long as it is represented by a continuous numerical range $[x_1, x_2]$ (Fig. \ref{fig:dofcorr}a). For a function defined on a two dimensional surface or multidimensional hyper-surface, continuity of domain of function is not a big issue when variables on different dimensions are independent, a two dimensional example was given with $x$ in the range $[x_1, x_2]$ and $y$ in the range $[y_1,y_2]$ (Fig. \ref{fig:dofcorr}b, the grey rectangular shadow indicates specified domain of a given function $f(x,y)$). However, for a functional defined on correlated variables, specifying range of respective degrees of freedom (DOFs) is no longer sufficient for defining domain of function as illustrated by Fig. \ref{fig:dofcorr}c, with grey-shadowed domain of function being a different region from the rectangle specified by ranges of $x$ and $y$, and understanding correlations between (among) involved variables (e.g. $x$ and $y$ here) becomes a necessity. More importantly, continuity of domain of function may start to be a concern as illustrated by the quasi-continuous scenario in Fig. \ref{fig:dofcorr}d and discrete regions in Fig. \ref{fig:dofcorr}e.  With increasing number of dimensions for a hyper-surface, complexity of topology for effective domain of a functional may rapidly become intractable. It is noted that for a multi-dimensional hyper-surface, fixing values of selected DOFs facing the similar problem. For example, one may simply imagine there is a third dimension in Fig. \ref{fig:dofcorr}d and Fig. \ref{fig:dofcorr}e with corresponding variable fixed. This is an important aspect of the curse of dimensionality for free energy calculation, and we attempt to bring it to the attention of the community. 

Free energy for a macrostate $M$ specified by $m$ out of $n$ ($1 \le m \le n$) degrees of freedom (DOFs) of a molecular system is a functional:
\begin{equation}
F_{M} = F([r_1], [r_2], \cdots, [r_m])\label {eq:FeTot}
\end{equation}
here we use square brackets to indicate that each of selected DOF may have a fixed value or vary within given range(s). It is noted that when all DOFs of a molecular system are fixed, we have a well-defined microstate and only energy matters in such cases since entropy for a given microstate is zero. While we may symbolically specify a macrostate with value or range of selected DOF, it is important to note that thermally accessible range of any given DOF is determined by values of all molecular DOFs that interact with it. With no further information, we may safely write that each DOF interacts with all other DOFs as the following:  
\begin{align}
%r_1 &= R_1(r_2, r_3, \cdots, r_n)\\
%\vdots\nonumber \\
[r_i] = R_i(r_1, r_2, \cdots, r_{i-1}, r_{i+1}, \cdots, r_n), (i = 1, 2, \cdots, m)
%\vdots\nonumber \\
%r_n & = R_n(r_1, r_2, \cdots, r_{n-1})
\end{align}
When one takes it for granted that a macrostate specified by fixed values and ranges of selected DOFs is continuous in configurational space, it is implied that all possible combinations of values within specified range of DOFs are practically thermally accessible, which is not necessarily true. Therefore, it is immediately clear that to ensure understanding continuity of an given region specified by ranges or fixed values of one or more DOFs in such a hyper-surface is a very challenging task, which implicates understanding of the joint distribution $P(r_1, r_2, \cdots, r_n)$ (i.e. the local free energy landscape(FEL)) in given patch(es) of the configurational space. The joint distribution(s) are part of goals rather than start point of our research on any molecular systems. To calculate free energy difference in a straight forward way, we first need to clearly define both end macrostates and therefore are in a dilemma of needing part of what we are looking for (correlations of the underlying variables within each of two end macrostates, which is essentially the excess (non-ideal gas) entropy). When both end macrostates are relatively easy to sample by brute force molecular dynamics (MD) or Monte Carlo (MC) simulations, various forms of principal component analysis (PCA) of trajectories can be effectively utilized to generate information on both free energy difference and corresponding domain in configurational space\cite{Mu2005,Zhou2006,Zhuravlev2010}. Unfortunately, many large scale conformational change of proteins and protein complexes occur on milli-second or longer time scales, which is expected to remain difficult to achieve on a routine basis in the near future. Therefore, more involved free energy methods are necessary. 

%We usually have very limited understanding for correlations among different DOFs for a complex molecular system. 

Historically, free energy calculation has been mainly carried out for two relatively well defined end macrostates with representative structural states for both end macrostates being available from high resolution experimental methods (e.g. X-ray diffraction)\cite{Monzon2013,Hrabe2015}. 
%Other methodologies that integrate out motion of fast DOFs include metadynamics\cite{} and essential dynamics\cite{}, these methods effectively reduces dimensions as indicated in equations \ref{eq:FeTot}. Simpler methods, such as various forms of elastic network models\cite{}, select to directly neglect large numbers of molecular DOFs. 
Nonetheless, since no information of relevant DOF correlations are available, we are limited to specifying end macrostates with ranges or fixed values of selected DOFs, and complications may occur. Apparently, thermally accessible domain in configurational space for free energy of a complex molecular system may potentially be much more complex than what illustrated in Fig. \ref{fig:dofcorr}. However, for convenience of illustration, we still utilize these simple scenarios in two dimensions as examples. A well behaved continuous macrostate is shown in Fig. \ref{fig:dofcorr}c, where specifying range of two dimensions is sufficient to define an easy-to-sample physical macrostate of corresponding system. In the quasi-continuous scenario in Fig. \ref{fig:dofcorr}d, one or more well-behaved substates in the specified subspace may be of practical relevance for a complex molecular system, and both situations present difficulty for free energy calculation. The discrete scenario in Fig. \ref{fig:dofcorr}e seems to contradict our widely utilized ergodic assumption, it is actually of great relevance for biomolecular systems. One typical type of examples are ligands that bind to a target protein in multiple independent and mutually non-transformable modes\cite{Mobley2009,Chodera2011}. More specifically, such ligands have to disassociate from its target protein before adopt an alternative pose of binding. We discuss specific issues of present mainstream free energy methods for the quasi-continuous and discrete scenarios in more details below.     
%This is not the case anymore starting from two interacting dimensions (Fig. 1b,c) and the complexity of figuring out continuity quickly explodes with increasingly large dimensions. While the curse of dimensionality is widely recognized, it usually refers to prohibitive computational efforts needed in multidimensional integration, not the difficult. Analysis of physical molecular systems is one typical example where continuity in high dimensional space is an important issue, which we elaborate below.

%The ability to understand, predict, manipulate and design various molecular systems is of great importance in physical, chemical, biomedical and nanotechnology research and industries. The core function(al) is free energy and great efforts have been invested to develop methodologies to calculate or estimate change of free energy for given molecular processes\cite{Series}. Since free energy for complex molecular systems is defined in high dimension, continuity is conceivably an inevitable difficult topic need to be addressed. However, systematic discussion of which is yet to be done except for a few special scenarios. I attempt to bring up this issue in a more general sense, and hopefully stimulate further interests in this regard and facilitate development of more efficient and reliable next generation free energy methodologies.
\section{Macrostate (or path) continuity for theoretically rigorous methodologies}
Thermodynamic integration (TI)\cite{Kirkwood1935,Darve2001} is a well-established methodology for calculation of free energy difference between two end macrostates of a given molecular system. Two fundamental principles underly this type of methods. The first is that change of free energy between two macrostates may be expressed as integration of the ensemble averaged energy over a path connecting them. The second is that free energy is a state function and the resulting integral is in principle not dependent upon specific path one choose, therefore one may choose unphysical integration paths based on operational convenience. In reality, despite this freedom of choosing integration path, one has to face the fact that the path is on a high dimensional hyper space, assure continuity and sufficient smoothness for reducing integration error, is fundamentally difficult (see Fig. \ref{fig:pathcont}). Alchemical transformation is widely utilized in thermodynamic integration due to its operational convenience, one importance issue is the well-understood ``end point catastrophe'', the singularity is caused essentially by insertion of a new correlated dimension in the original hyper-surface. This issue has been successfully addressed by utilization of soft core potentials\cite{Steinbrecher2011}.  Another major challenge of TI is potential existence of slower DOFs in addition to selected integration path, which may cause convergence difficulty for the calculation if insufficient sampling of such DOFs changes statistically significant configurational subspace of either or both end states. One accompanying problem is that insufficient sampling of such slower DOFs (with dynamics significantly slower than that of integration path) may also influence continuity of both end macrostates in unpredictable way. %, the configurational continuity corresponds to this well acknowledged scenario is schematically shown in Fig. XXX. 

TI formulations, by utilizing one specific single integration path, require that both end macrostates are continuously-defined region on the relevant high dimensional free energy landscapes. Apparently, when one of the end macrostate specified by fixed values or ranges of various DOFs are quasi-continuous or discrete as shown in Fig. \ref{fig:dofcorr}de, TI is problematic and may fail in different ways: i) all quasi-continuous ( or discrete ) regions are functionally important but only one is sampled, resulting in underestimation of statistical weight; ii) the sampled region is different from functionally relevant region but fall in the same ``box'' as specified by the range of relevant DOFs, both underestimation and overestimation of statistical weight for this end macrostate is possible. In the case that one or both end macrostates are comprising quasi-continuous or discrete substates, the difficulty of ensuring path continuity increases, especially near such end macrostates.
In histogram based methods, such as various forms of umbrella sampling\cite{} or weighted histogram analysis\cite{}, a reaction coordinate is selected to connect the start and final macrostates, both of which are specified by ranges or values of selected molecular DOFs. Therefore, as in the case of TI, it is necessary to be careful that proper physical states are sampled in trajectories to ensure correct calculation of free energy differences. 

%This assumption may easily be violated. There are scenarios that one or both end macrostates are fundamentally discrete regions on FEL. For example, if a ligand has two distinct modes of binding on a protein, and transitions between these two modes may not be realized without dessociation of the ligand from the protein molecule, then the ``bound state'' is actually two discontinuous region. Apparently, in such scenarios, it is not possible to select one smooth continuous path connecting the end macrostates any more. Even for well-defined continuous region on FEL, difficulties may rise in computation. In computation, we usually utilize range of selected DOFs to define macrostates, and such practices may cause significant error. A specific scenario is presented in Fig. 2, where a two dimensional end state (shadowed ellipsoid) is well-defined and enclosed in the region specified by the rectangle. However, there is another irrelevant region (open ellipsoid) in the same rectangle. When we do not understand detailed correlation of selected DOFs within the range, which is true in most cases for high dimensional problems, that irrelevant state may be taken as our end macrostate of part of it. Such possibilities rapidly increase with increasingly high dimensions. Assuming infinite computational resources, eventual convergence of slower DOFs may expand one or both end macrostates, or possibly reach quasi-discrete regions. The above mentioned issues exist similarly for weighted histogram analysis and other similar methods.  
     
In yet another path-based rigorous free energy calculation method, namely non-equilibrium work (NEW) based analysis\cite{Jarzynski1997,Hummer2001,Goette2008}. Situations change slightly. Since in this method, numerous paths are generated to realize non-equilibrium transitions between the two end macrostates, the absolute continuity is no-longer required for the final macrostate, as sampling of discrete regions in the final macrostate is realizable, at least in theory, when sufficient number of non-equilibrium paths are sampled. However, when trajectories are utilized to sampling the start macrostate, an quasi-continuous or discrete scenario present apparent challenge for sampling. The fundamental difference between integration path in TI and transition paths in NEW is that the former may be any unphysical ones that are convenient for calculation, while the later are physical paths since actual work done along which are exponentially averaged to estimate change of free energy as shown below\cite{Jarzynski1997}:
\begin{equation}
\overline{exp(-\beta W)} = exp(-\beta \Delta F)
\end{equation}
When the final macrostate has quasi-continuous or discrete substates, relative dynamical accessibility of various substates determines difficulty of convergence for calculated free energy difference. When dynamical accessibility of multiple substates are proportional to their respective statistical weights, convergence may be achieved with relative ease. On the contrary, when heavier substates is more difficult to access dynamically, convergence of NEW methods will deteriorate and significantly more transition processes between two end macrostates need to be recorded.  

Free energy perturbation method (FEP)\cite{Zwanzig1954,Zwanzig1955,Bash1987} was originally proposed as an end point method with no need of designing reaction coordinate (or integration pathway). Reliable calculation of free energy difference by FEP requires sufficient overlapping of statistically significant region of configurational space between two end macrostates. When one or both macrostates exhibit quasi-continuous or discrete domain in configurational space, overlapping between two end macrostates is apparently more difficult to achieve when compared with situations that both end macrostates have well-behaved continuity as illustrated in Fig. \ref{fig:dofcorr}c. When one or both end macrostate(s) exist as discrete or quasi-continuous regions in FEL with \emph{a priori} unknown correlation of relevant DOFs, multistage FEP\cite{MFEP1999}, which is designed to alleviate insufficient overlapping of statistically significant configurational space of end macrostates via utilization of an effective order parameter, face similar challenges as TI and histogram based methodologies do.

Hypothetical scanning molecular dynamics (Monte Carlo) (HSMD/MC)\cite{Meirovitch1999,White2004,Cheluvaraja2005,Cheluvaraja2008,Meirovitch2010} is a recent rigorous end point free energy method based on reconstruction of transition probability to specific configurations generated from regular MD or MC sampling. As reconstruction process is limited to very small configurational space volume in practice, the caveat of including irrelevant region is not a big issue in this case. The potential problem would be from the sufficiency of sampling provided by the starting molecular configurations, which may neglect some discrete or quasi-discrete substates that are of relevance in our interested molecular processes, or alternatively include some discrete or quasi-continuous substates that are essentially irrelevant. Both ways lead to unreliable results.

These rigorous methods usually are utilized to calculate free energy change between end macrostates that are quite well defined structurally (i.e. with structures of both end macrostates available), with the goal of providing atomistic explanations for experimental observations. In these situations, while there is possibility of a functionally relevant state comprising a limited number of quasi-continuous or discrete regions in configurational space, functional robustness and evolution pressure excludes possibility of existence of a large number of statistically significant quasi-continuous or discrete regions. Nonetheless, when starting with structures of both end macrostates, no correlation information is available and care has to be taken for possible configurational discontinuity. 
Historically, while continuity of macrostate has not been articulated, a number of dimensionality reduction methods have been developed to find the genuine reaction path between interested macrostates\cite{}. As a matter of fact, these methodologies may be utilized to detect potential continuity problem of any give macrostate, albeit with high computational costs. For example, a \emph{posterio} principal component analysis (PCA) of configurational space visited by sampling trajectories in the vicinity of experimental structures, while unable to reveal possible unvisited statistically significant quasi-continuous or discrete regions, should at least disclose the topology and continuity of the visited configurational subspace. One may also using metadynamics approach\cite{} to probe the free energy landscape of the interested molecular systems in the vicinity of both end macrostates to provide guidance for selecting proper rigorous free energy methodology. The most challenging part of metadynamics is to select proper collective variables (CVs), which should be the most critical slow DOFs in molecular systems. A number of strategies\cite{} are available to facilitate definition of CVs. Nonetheless, defining proper CVs is a system specific task and remains to be a major challenge to be tackled. 

\section{Macrostate continuity and major approximate methodologies}
Due to the prohibitive computational cost, the above mentioned rigorous methods are not practical for high throughput predictive calculation of free energies as in the case of protein folding, design, docking and virtual screening. A number of computationally more economical methods\cite{LIE1998,MMPBSA-JMC,Miller2012} and numerous scoring functions\cite{Grinter2014,Liu2015} have been developed over last few decades. We discuss the relevance of configurational space continuity for them below.

In linear interaction energy (LIE) model\cite{LIE1998}, change of free energy upon the binding of a ligand ($L$) to its target protein ($P$) is estimated by simulating the free ligand in solution and the protein-ligand complex ($PL$) with the following equation:
\begin{equation}
\Delta G = \beta(\langle E^{L-S}_{ele}\rangle_{PL} - \langle E^{L-S}_{ele}\rangle_L) + \alpha(\langle E^{L-S}_{vdW}\rangle_{PL} - \langle E^{L-S}_{vdW}\rangle_L)
\end{equation}
with $E^{L-S}_{ele}$ and $E^{L-S}_vdW$ being the electrostatic and van der Waals interaction energies between the ligand and it environment ($S$, including protein and solution), angle brackets indicating ensemble average, and $\alpha$ and $\beta$ are two empirical parameters. For a rather rigid protein-ligand binding, where both free and bound state are likely to be well behaved and continuous in configurational space as illustrated in Fig. \ref{fig:dofcorr}c, the approximation would be good given proper parameters $\alpha$ and $\beta$. However, when significant flexibility exist for ligand and/or its target, either free and bound state may have more than one quasi-continuous or discrete substates as illustrated in Fig. \ref{fig:dofcorr}d or Fig. \ref{fig:dofcorr}e, one set of parameter is likely to fail on some of such substates and reliability of the approximation starts to deteriorate. 

The linear response approximation (LRA)\cite{Sham2000} is quite similar to LIE. The practical difference between them is that non-electrostatic contribution for the former is evaluated with protein dipoles langevin dipoles (PDLD) method (or its semi-microscopic version, the PDLD/s method)\cite{Singh2009} while for the later is approximated by averaging van der Waals interactions. Nonetheless, both treatments assume that end macrostates are well-defined free energy wells as illustrated by Fig. \ref{fig:dofcorr}c. The restraint-release method\cite{Singh2009}, which decomposes conformational change into three contributions, energy difference between two representative (central) structures of the start and final conformation, entropic contribution from local motion around the central structure in start conformation, and entropic contribution from local motion around the central structure in the final conformation, implies the same assumption. One may imagine that it would be highly nontrivial, if ever possible, to pick a ``representative (central)'' structure for quasi-continuous or discrete scenarios. 

In MM/P(G)BSA\cite{MMPBSA-JMC}, free energy of protein-ligand (in a general sense) binding is expressed as:
\begin{align}
\Delta G &= G(PL) - G(P) - G(L) \\
G &= \langle E_{int} + E_{ele} + E_{vdW} + G_{solv} + G_{np} - TS_{MM}\rangle \label{eq:gpart}
\end{align}
the three first terms in equation (\ref{eq:gpart}) are the molecular mechanical ($MM$) internal, electrostatics and van der Waals energies, $G_{solv}$ and $G_{np}$ are polar and non-polar solvation free energies respectively. $S_{MM}$ is the configurational entropy estimated with normal mode\cite{}or quasiharmonic analysis\cite{}. Again, the validity of this equation implies that both end macrostates are well-behaved as illustrated in Fig. 1c. Apparently, both normal mode and quasiharmonic analysis are insufficient methods for entropy estimation when a macrostate have quasi-continuous or discrete substates, estimation of solvation terms become significantly more challenging for these scenarios as well. 

Flexibility and rugged FEL is essentially a ubiquitous property and widely acknowledged challenge in computational analysis of protein molecular systems\cite{Panjkovich2012,Hrabe2015}. Entropic effects of ligands are also demonstrated to be important\cite{Villa2000}. Therefore, the implicit assumption by these approximate methods that concerned end macrostates have well-behaved continuity in configurational space is not necessarily true, especially for important and/or interesting molecular systems that we have very limited understanding and therefore have strong desire to predict their behavior. In high throughput predictive studies, we usually seek to locate statistically significant configurational subspace by some predesigned procedures. In contrast to application of rigorous methods in comparing macrostates which are already established to be statistically significant, scanning of configurational space faces much greater challenge of continuity. Since we know very little on time scales and correlations of relevant moelcular DOFs, we may well specify ``a'' macrostate that is physically comprising many quasi-continuous or even discrete regions in configurational space, estimation of collective statistical weight of which would be highly unreliable when utilizing approximate methods 
that require well-defined continuous end macrostates. In a recent analysis on the utility of various minimum potential terms (minimum protein self energy, minimum protein-solvent interaction energy and their sum) as approximate free energy proxy, correlation of these terms with population based free energy differences deteriorate dramatically when macrostates were defined by projection onto fast torsional DOFs, which essentially correspond to defining many discrete fragments in configurational space as a collective macrostate. Methods that are based on the assumption of well defined continuous macrostates in configurational space will lose their prediction power in such situation.

As discussed above, the effort of free energy calculation has been predominantly focusing on, understandably, the difference between two end macrostates or paths connecting them when using path-based methods. Nonetheless, configurational space continuity within each end macrostate is just as important in terms of reliability of calculation/estimation. When one takes configurational space continuity of ``a'' macrostate specified by ranges or values of selected DOFs for granted, unexpected errors caused by quasi-continous or discrete configurational substates may severely reduce reliability of calculation.

%\section{Scoring functions and configurational space continuity} 
The most widely utilized free energy methods in high throughput applications (e.g. virtual screening) are various forms of scoring functions\cite{}, parameters for which are derived based upon presently available structures and/or affinity data. Regardless of specific type, scoring functions usually evaluate, compare and search for the best ``pose'' of molecular interactions. Let's assume that a given scoring function is sufficiently accurate, the precondition of finding the best ``pose'' is to have it in a limited list of ``poses'' to be evaluated. Preparing such a list is an extremely challenging global sampling problem in the configurational space. The key issue here is that we have no idea which specific region(s) of configurational space is represented by a given ``pose'', which is essentially a point (or a microscopic patch) in a gigantic multidimensional configurational hyper-surface. Consequently, hierarchical structure of FEL for complex molecular systems (e.g. a protein-ligand interaction system) may not be effectively utilized to accelerate the searching process.
Therefore, the capability to measure the statistical weight of an arbitrarily given configurational subspace ( regardless of its actual physical continuity ) is highly desired, a schematic representation of hierarchical configurational space partition is presented in Fig. \ref{fig:???}. Achieving this goal makes guaranteed sampling of the global configurational space possible through proper partitions into subspaces. ????

\section{Calculation of free energy difference by direct configurational space discretization}
We recently developed an end point free energy method based on direct discretization of configurational space into explicit conformers\cite{Wang2016}. More specifically, one first need to define a set of Explicit Conformers with Invariable Statistical Weight Distribution (ECISWD) across the whole configurational space for the interested molecular system, and free energy difference between two end macrostates may be obtained simply by counting thermally accessible such conformers as shown in the following equation:
\begin{equation}
\Delta F^{AB} = -k_BTln\frac{N^A_{conf}}{N^B_{conf}}
\end{equation}
with $k_B$ being the Boltzmann constant, $T$ being the temperature, and $N^{A(B)}_{conf}$ being the number of ECISWD in the two end macrostates $A$ and $B$ respectively. By combining sequential Monte Carlo (SMC) and importance sampling\cite{Zhang2003,Zhang2006}, counting of conformers may be achieved highly efficiently. In this methodology, continuity of configurational space is not an implied assumption for end macrostates to be compared. However, in importance sampling, each replica apparently may only represent one specific discrete region when many exist. Therefore, while physical continuity is not implied, increase of the number of discrete or quasi-discrete regions in one specified ``macrostate'' might present challenges and requires larger number of replica for sampling. This is not necessarily the case as when indeed many discrete regions were included in an artificial macrostate, sampling only part of them may provide us with sufficient accuracy already. More investigations are therefore necessary in this regard. While not limited by configurational space continuity assumption, configurational space discretization based free energy method inherently does not ensure continuity for the identified configurational subspace that is the most statistically significant, and one or multiple conformations may be included. Again, metadynamics approach may be utilized to probe the details of FEL within ``a'' statistically significant region, which may physically correspond to one or a number of quasi-continuous or discrete regions. Development of more efficient methods for identifying possible configurational space (quasi-)discontinuity in ``a'' specified ``macrostate'' is highly desired for more effective utility of all present free energy methodologies.    

\section{Conclusions}
Free energy calculation/estimation is of long-standing interest in computational chemistry, molecular biophysics and biology. There have been numerous excellent original articles, reviews and books on this subject and we apologize for not being able to citing them all here. Readers are encouraged to read more extensively for details of each type of methods. In this perspective, we limited our discussions to challenges of configurational space continuity facing the free energy calculation community. While simple low dimensional schematics were utilized for explanation, real high dimensional molecular systems present similar but conceivably much more complex scenarios. 

We illustrate that either explicit consideration for detecting configurational space continuity problem or development of new methods that do not require configurational space continuity of end macrostates ( and/or pathway continuity ) need to be addressed in theoretical and computational advancement of free energy analysis.

%%%%%%%%%%%%%%%%%%%%%%%%%%%%%%%%%%%%%%%%%%%%%%%%%%%%%%%%%%%%%%%%%%%%%
%% The "Acknowledgement" section can be given in all manuscript
%% classes.  This should be given within the "acknowledgement"
%% environment, which will make the correct section or running title.
%%%%%%%%%%%%%%%%%%%%%%%%%%%%%%%%%%%%%%%%%%%%%%%%%%%%%%%%%%%%%%%%%%%%%
\begin{acknowledgement}
This research was supported by National Natural Science Foundation of China under grant number 31270758, and by the Research fund for the doctoral program of higher education under grant number 20120061110019. %Computational resources were partially supported by High Performance Computing Center of Jilin University, China. 
\end{acknowledgement}

%%%%%%%%%%%%%%%%%%%%%%%%%%%%%%%%%%%%%%%%%%%%%%%%%%%%%%%%%%%%%%%%%%%%%
%% The same is true for Supporting Information, which should use the
%% suppinfo environment.
%%%%%%%%%%%%%%%%%%%%%%%%%%%%%%%%%%%%%%%%%%%%%%%%%%%%%%%%%%%%%%%%%%%%%
%\begin{suppinfo}
%
%A listing of the contents of each file supplied as Supporting Information
%should be included. For instructions on what should be included in the
%Supporting Information as well as how to prepare this material for
%publications, refer to the journal's Instructions for Authors.
%
%The following files are available free of charge.
%\begin{itemize}
%  \item Filename: brief description
%  \item Filename: brief description
%\end{itemize}

%\end{suppinfo}

%%%%%%%%%%%%%%%%%%%%%%%%%%%%%%%%%%%%%%%%%%%%%%%%%%%%%%%%%%%%%%%%%%%%%
%% The appropriate \bibliography command should be placed here.
%% Notice that the class file automatically sets \bibliographystyle
%% and also names the section correctly.
%%%%%%%%%%%%%%%%%%%%%%%%%%%%%%%%%%%%%%%%%%%%%%%%%%%%%%%%%%%%%%%%%%%%%
%\bibliography{../Total2,./corr,../Flexibility}
\providecommand{\latin}[1]{#1}
\providecommand*\mcitethebibliography{\thebibliography}
\csname @ifundefined\endcsname{endmcitethebibliography}
  {\let\endmcitethebibliography\endthebibliography}{}

\newpage

\begin{figure}
\centering 
{\includegraphics[width=4.5in]{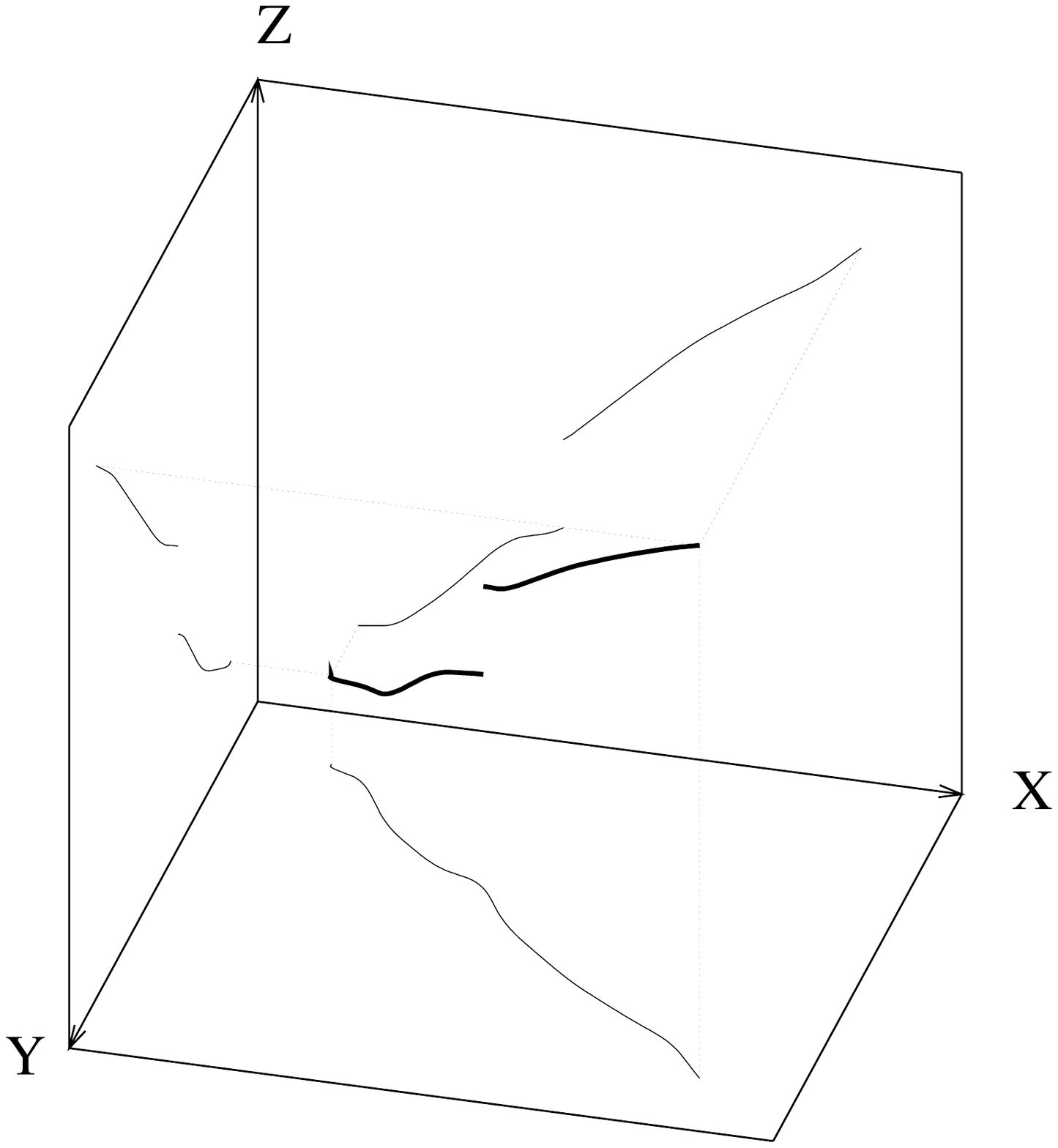}}
\caption{Schematic representation of path continuity in three dimensions. This path is discontinuous in three dimensional space but have a continuous project in one of the plane ($x$,$y$)} 
\label{fig:pathcont}
\end{figure}

\begin{figure}
\centering 
\includegraphics[width=6in]{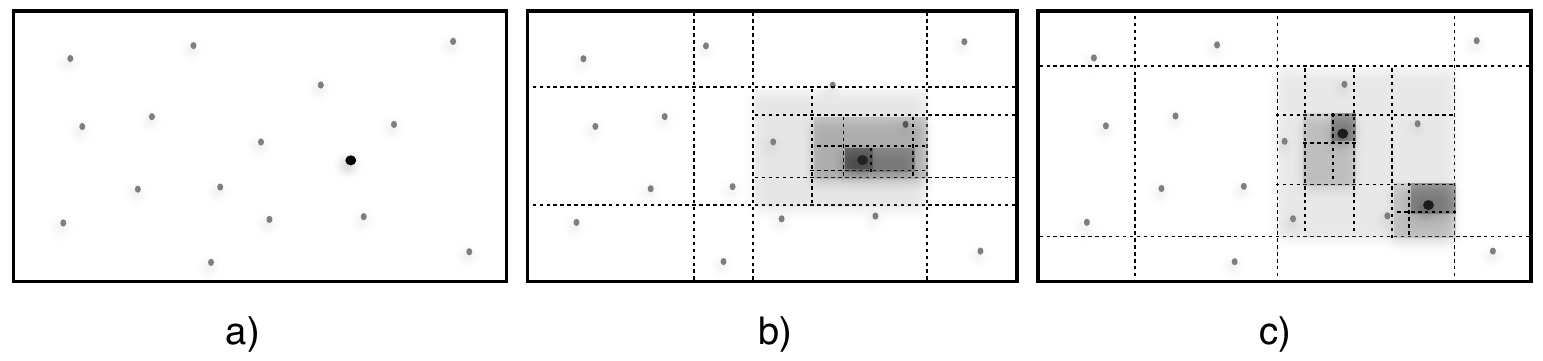}
\caption{Schematic representation of a searching process for the most statistically significant macrostate(s). The rectangle represents the full configurational space of a given molecular system, large dark dots represent the lowest free energy state(s), the small grey dots represent other poses to be examined, it is important to note that size of these dots and the rectangle are not to the proportion. As a matter of fact, the configurational space, the size of which grows exponentially with number of DOFs and is huge for typical biomolecular systems, was intentionally given an extremely small size for the convenience of representation. a) ``pose'' strategy where many poses were examined and the one with lowest free energy (as indicated by the selected scoring function) is deemed as the answer, the precondition of finding the right answer is that the answer is already in the list of poses to be evaluated. b) A hierarchical strategy to locate for the statistically most significant configurational subspace. On each hierarchy, the most significant subspace is given a darker background that other subspaces, there are four hierarchies in this particular case. c) A configurational space with two statistically most significant states were searched with a hierarchical strategy.}
\label{fig:pathcont}
\end{figure}

\end{document}